\documentclass[prb,aps,twocolumn,showpacs,dvips]{revtex4}
\usepackage{feynmp}
\usepackage{amssymb}
\usepackage{epsfig}
\usepackage{graphicx}
\usepackage{subfigure}
\usepackage{hyperref}

\begin{document}

\title{Fate of non-Fermi liquid behavior in QED$_{3}$ at finite chemical potential}
\author{Jing-Rong Wang and Guo-Zhu Liu \\
{\small {\it Department of Modern Physics, University of Science and
Technology of China, Hefei, Anhui, 230026, P.R. China }}}

\begin{abstract}
The damping rate of two-dimensional massless Dirac fermions exhibit
non-Fermi liquid behavior, $\propto \varepsilon^{1/2}$, due to gauge
field at zero temperature and zero chemical potential. We study the
fate of this behavior at finite chemical potential. We fist
calculate explicitly the temporal and spatial components of vacuum
polarization functions. The analytical expressions imply that the
temporal component of gauge field develops a static screening length
at finite chemical potential while the transverse component remains
long-ranged owing to gauge invariance. We then calculate the fermion
damping rate and show that the temporal gauge field leads to normal
Fermi liquid behavior but the transverse gauge field leads to
non-Fermi liquid behavior $\propto \varepsilon^{2/3}$ at zero
temperature. This energy-dependence is more regular than $\propto
\varepsilon^{1/2}$ and does not change as chemical potential varies.
\end{abstract}


\maketitle


\section{Introduction}

The damping rate of fermions due to interaction with gauge fields is
a physical quantity of broad interests. Studying this quantity can
help us to judge whether an interacting fermion system displays
non-Fermi liquid behavior or not. According to Landau, for any
normal Fermi liquid to be stable, the fermion excitations must have
a sufficiently long lifetime, which means that the fermion damping
rate should vanish faster than energy $\varepsilon$ does as
$\varepsilon \rightarrow 0$. At the low-energy regime, the fermion
damping rate can be written in the form
$\mathrm{Im}\Sigma(\varepsilon)\propto \varepsilon^z$. The system
with exponent $z > 1$ is a normal Fermi liquid, while system with $z
\leq 1$ corresponds to a non-Fermi liquid. In conventional metals,
the Coulomb interaction between electrons is always statically
screened and can only lead to normal Fermi liquid behavior. Since
the work of Holstein and coworkers \cite{Holstein}, it has been
known that the unscreened gauge field can give rise to non-Fermi
liquid behavior. The unusual, non-Fermi liquid like, fermion damping
rate caused by gauge field has attracted great attention in the past
twenty years because an emergent gauge field is found to play
important roles in a number of strongly correlated electron systems
\cite{Varma2002, Reizer89, Lee89, Blok93, GanWong, Polchinski,
Lee05}. In particle physics, this problem has also been discussed
extensively in various gauge theories, including four-dimensional
QCD \cite{Pisarski} and four-dimensional QED \cite{Lebellac,
Blaizot}.

Here we are particularly interested in the unusual properties of
massless Dirac fermions. In some planar correlated electron systems,
including $d$-wave high temperature superconductor \cite{Lee05} and
graphene \cite{Castro}, the valence band and conduction band touch
only at discrete Dirac points. This is illustrated in Fig.1(a). The
states in the lower valence band are fully occupied, while those in
the upper conduction band are fully empty. The low-energy fermions
excited from the lower band have a linear spectrum and hence can be
described by massless Dirac fermions, which satisfy the relativistic
Dirac equation \cite{Lee05, Castro}. The interaction of massless
Dirac fermions with an abelian gauge field in two spatial dimensions
defines the three-dimensional quantum electrodynamics (QED$_{3}$).
This field theory is usually studied by particle physicists as a toy
model of QCD$_4$ since it is known to exhibit dynamical chiral
symmetry breaking \cite{Appelquist88} and confinement
\cite{Maris95}. In condensed matter physics, with proper
modification, it serves as an effective low-energy theory of high
temperature superconductors \cite{Affleck, Kim97, Franz, Kim99} and
some spin-1/2 Kagome spin liquids \cite{Ran07}. In the realistic
applications, chiral symmetry breaking and confinement in QED$_{3}$
correspond to the long-range Neel order in two-dimensional quantum
Heisenberg antiferromagnet \cite{Kim99}.

Recently, we studied the damping rate of massless Dirac fermions due
to gauge field in QED$_3$ in the absence of chiral symmetry breaking
and confinement \cite{WangLiu}. It diverges at both zero and finite
temperatures when it is calculated using the straightforward
perturbative expansion approach. Once the fermion damping effect and
the dynamical screening effect of gauge field are self-consistently
coupled, the fermion damping rate is then well-defined and behaves
as $\mathrm{Im}\Sigma(\varepsilon) \propto \varepsilon^{1/2}$ at
zero temperature \cite{WangLiu}. This damping rate vanishes slower
than $\varepsilon$ does in the $\varepsilon \rightarrow 0$ limit and
thus displays non-Fermi liquid behavior.

\begin{figure}[ht]
  \centering
    \subfigure{
    \includegraphics[width=1.6in, bb=15 15 235 235, clip, angle=0]{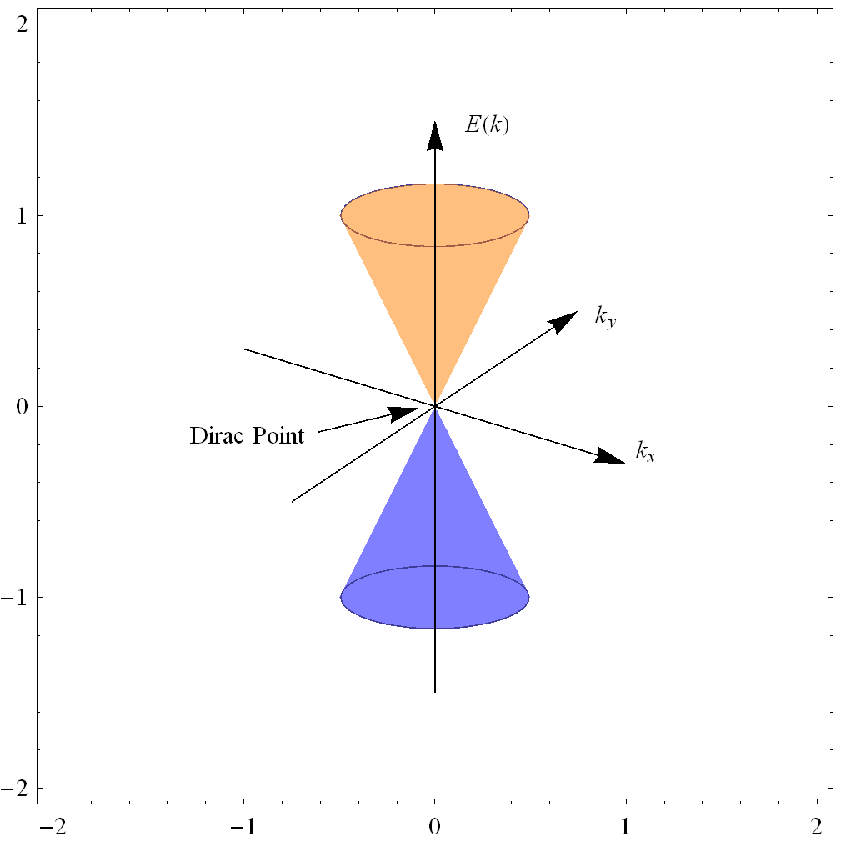}}
    \subfigure{
    \includegraphics[width=1.6in, bb=15 15 235 235, clip, angle=0]{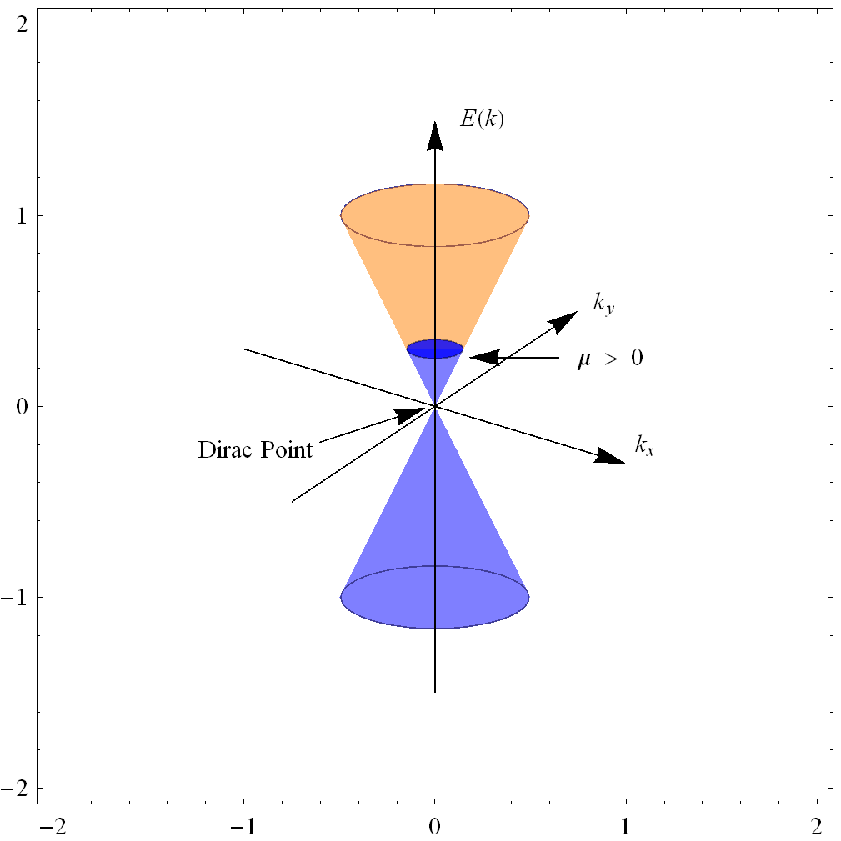}}
\caption{(a) The half-filling state with lower band being fully
occupied and upper band fully empty. (b) At finite chemical
potential $\mu$, the Fermi surface is finite.}
\end{figure}

This result was obtained in the so-called half-filling ground state
depicted in Fig.1(a). In realistic condensed matter systems, the
fermion density already can be continuously turned, either by
chemical doping \cite{Lee05} or by adjusting a bias voltage
\cite{Castro}. When the fermion density grows from the Dirac point,
a small but finite Fermi surface will emerge, as shown in Fig.1(b).
To describe this process, the commonly used strategy is to introduce
a chemical potential $\mu$, which defines the energy difference
between the new Fermi surface and the Dirac point. The systems at
zero and finite $\mu$ may have quite different properties. Indeed,
there might be a quantum phase transition as the chemical potential
varies \cite{Sachdev}, with $\mu = 0$ being the quantum critical
point.

When the fermion density becomes sufficiently large, the interacting
fermion system finally develops a large Fermi surface with
low-energy fermionic excitations satisfying the non-relativistic
Schrodinger equation. Historically, before the energy gap of high
temperature cuprate superconductor was confirmed to have a $d$-wave
symmetry, the fermion-gauge system with large Fermi surface had been
studied intensively \cite{Varma2002, Lee89, Blok93, GanWong,
Polchinski, Lee05}. The damping rate of non-relativistic fermions
was calculated by various methods, including straightforward
perturbation expansion \cite{Lee89, Blok93}, renormalization group
approach \cite{GanWong}, and Eliashberg theory \cite{Polchinski}.
Most of these studies found that $\mathrm{Im}\Sigma(\varepsilon)
\propto \varepsilon^{2/3}$ at zero temperature. The exponent in the
energy dependence of damping rate is quite different from that in
QED$_3$ of massless Dirac fermions at $\mu = 0$. The difference is
presumably owing to the difference between a large Fermi surface and
discrete Fermi points.

As $\mu$ grows from $\mu = 0$ to a large value, the fermion damping
rate will undergo a crossover from $\propto \varepsilon^{1/2}$ to
$\propto \varepsilon^{2/3}$. A question naturally arises: Does the
exponent $z$ appearing in $\varepsilon^{z}$ vary continuously from
$1/2$ to $2/3$ or change abruptly at some critical value $\mu_c$?

In this paper, we study how the fermion damping rate varies with
growing chemical potential $\mu$ by calculating the $\mu$-dependence
of fermion self-energy. As usual, the gauge field is decoupled to
longitudinal and transverse components. From the vacuum polarization
functions at finite $\mu$, we know that the longitudinal component
becomes short-ranged (massive) due to static Debye screening effect,
but the transverse part remains long-ranged (massless) because of
the gauge invariance. In this case, the transverse component of
gauge field dominates and should be able to produce non-Fermi liquid
behaviors. However, with increasing fermion density, the dynamical
screening effect becomes stronger and hence may lead to less
singular behavior than that at small $\mu$.

After explicit computation, we found that the transverse damping
rate of Dirac fermions at zero temperature behaves as
$\mathrm{Im}\Sigma_{\mathrm{T}}(\xi_{\mathbf{k}})\propto
\mu^{-1/3}\xi_{\mathbf{k}}^{2/3}$, while the longitudinal
contribution is $\mathrm{Im}\Sigma_{\mathrm{L}}(\xi_{\mathbf{k}})
\propto \left(\xi_{\mathbf{k}}^2/\mu\right)
\ln\left(\xi_{\mathbf{k}}/\mu\right)$, where $\xi_{\mathbf{k}}$ is
the fermion energy in the on-shell approximation. Thus the total
fermion damping rate is $\propto \mu^{-1/3}\xi_{\mathbf{k}}^{2/3}$,
which is certainly non-Fermi liquid behavior. We also considered the
fixed momentum approximation and obtained the same results, i.e.,
the fermion damping rate behaves as $\propto
\mu^{-1/3}\varepsilon^{2/3}$ at zero temperature.

These results imply that the fermion damping rate suddenly becomes
$\propto \varepsilon^{2/3}$ from $\propto \varepsilon^{1/2}$ once
the chemical potential $\mu$ departs from zero. As $\mu$ grows, the
energy-dependence of fermion damping rate does not change but its
coefficient decreases. Therefore, although the Dirac fermions are
always not well-defined in the sense of Landau quasiparticle, their
lifetime increases slowly with growing fermion density.

The paper is organized as follows. The Lagrangian and some relevant
quantities are defined In Sec. II. The full expressions of
polarization functions from massless Dirac fermions at finite
chemical potential are calculated in Sec. III and the fermion
damping rate is calculated in Sec. IV. We summarize the results and
briefly discuss the physical implications in Sec.V. The polarization
functions at zero temperature are shown in the Appendix.

\section{Lagrangian and Feynman rules of QED$_3$ at finite chemical potential}

We start from the following general Lagrangian of QED$_{3}$
\begin{equation}
\mathcal{L} = \sum_{i=1}^{N} \Psi_{i}^\dag\left(\partial_{\tau} -
\mu - ie a_{0} - i\mathbf{\sigma}\cdot \left(\mathbf{\partial}
-ie\mathbf{a}\right)\right)\Psi_{i} - \frac{1}{4}F^{2}.
\end{equation}
This is the general Lagrangian for QED defined in (2+1)-dimensional
space-time. As a well-defined relativistic quantum field theory,
there is surely an explicit Maxwell term for the gauge field. As
discussed in the Introduction, this is a very interesting and widely
studied field theory in the context of particle physics. When
applied to strongly correlated electron systems, it usually needs to
be modified properly. If the effective QED$_3$ theory is derived by
considering the phase fluctuations in underdoped high temperature
superconductors, then normally the Maxwell term is present
\cite{Franz}. However, if the effective QED$_{3}$ theory is obtained
by the slave-particle treatment of \emph{t}-\emph{J} model, there is
no Maxwell term in the Lagrangian and the gauge field can have its
dynamics only after integrating out the matter fields \cite{Affleck,
Kim97}. We will first consider the general action with the Maxwell
term and briefly discuss the case without such term at the end of
Sec.IV.

Here, we adopt the two-component representation of spinor field,
with $\mathbf{\sigma}$ being the Pauli matrices. The Dirac fermion
flavor $N$ is taken to be large so that we can use the $1/N$
expansion. The theory is defined at finite chemical potential $\mu$.
The aim of this paper is to study how the fermion damping rate
depends on $\mu$. For simplicity, we take $\hbar = c = k_B = 1$
throughout the whole paper.

At finite temperature, the Matsubara propagator of massless Dirac
fermion is
\begin{equation}
G_{0}\left(i\varepsilon_{n},\mathbf{k}\right) =
\frac{1}{i\varepsilon_{n}+\mu-\mathbf{\sigma}\cdot\mathbf{k}},
\end{equation}
where $\varepsilon_{n}=(2n+1)\pi T$ with $n$ being integer. After
analytic continuation, the retarded propagator is
\begin{equation}
G_{0}(\varepsilon,\mathbf{k}) =
\frac{1}{\varepsilon+\mu-\mathbf{\sigma}\cdot\mathbf{k}+i\delta}.
\end{equation}
At finite temperature, the temporal and spatial components of gauge field
decouple and now it is convenient to work in the Coulomb gauge
$k_{i}a_{i}=0$. In the imaginary time formalism, the propagator for
the gauge field can now be written as
\begin{eqnarray}
D_{00}\left(i\omega_{m},\mathbf{q}\right) &=&
\frac{1}{|\mathbf{q}|^2+\Pi_{00}(i\omega_{m},\mathbf{q})},\label{eqn:Photon00}
\\
D_{ij}\left(i\omega_{m},\mathbf{q}\right) &=& \left(\delta_{ij} -
\frac{q_{i}q_{j}}{\mathbf{q}^2}\right)\frac{1}{|\mathbf{q}|^2\
+\omega_{m}^{2}+\Pi_{\bot} \left(i\omega_{m},\mathbf{q}\right)},
\nonumber \\\label{eqn:Photonij}
\end{eqnarray}
where $\omega_{m} = 2m\pi T$ for bosonic modes with $m$ being
integers. The vacuum polarization functions
$\Pi_{00}(\omega_{m},\mathbf{q})$ and $\Pi_{\bot}(\omega_{m},\mathbf{q})$
come from the one-loop bubble diagram of Dirac fermions to the
leading order of $1/N$ expansion. In particular, the polarization
function appearing in the spatial component is given by
\begin{eqnarray}
\Pi_{\bot}(i\omega_{m},\mathbf{q}) =
\Pi_{ii}(i\omega_{m},\mathbf{q}) -
\frac{\omega_{m}^{2}}{\mathbf{q}^{2}}\Pi_{00}(i\omega_{m},\mathbf{q}).
\label{eqn:PolarizationBot}
\end{eqnarray}
The functions $\Pi_{00}(i\omega_{m},\mathbf{q})$ and
$\Pi_{ii}(i\omega_{m},\mathbf{q})$ are defined as
\begin{eqnarray}
\Pi_{00}(i\omega_{m},\mathbf{q}) &=& -Ne^2T\sum_{i\varepsilon_{n}}
\int\frac{d^{2}k}{(2\pi)^{2}}\nonumber
\\
&&\times\mathrm{Tr}[G_{0}(i\varepsilon_{n},\mathbf{k})
G_{0}(i\varepsilon_{n}+i\omega_{m},\mathbf{q}+\mathbf{k})], \\
\Pi_{ii}(i\omega_{m},\mathbf{q}) &=& Ne^2T\sum_{i\varepsilon_{n}}
\int\frac{d^{2}k}{(2\pi)^{2}}\nonumber
\\
&&\times\mathrm{Tr}[\sigma_{i}
G_{0}(i\varepsilon_{n},\mathbf{k})\sigma_{i}
G_{0}(i\varepsilon_{n}+i\omega_{m},\mathbf{q}+\mathbf{k})].
\nonumber \\
\end{eqnarray}
When we calculate the fermion damping rate, we need the real and
imaginary parts of the retarded polarization functions. They can be
obtained by straightforward computation, which will be given in the
next section.

The fermion damping rate can be calculated by the standard finite
temperature field theory technique \cite{Giuliani}. To the lowest
order of $1/N$ expansion, the self-energy of Dirac fermion is given
by
\begin{equation}
\Sigma(i\varepsilon_n,\mathbf{k})
=\Sigma_{\mathrm{L}}(i\varepsilon_{n},\mathbf{k}) +
\Sigma_{\mathrm{T}}(i\varepsilon_{n},\mathbf{k}),
\end{equation}
where
\begin{eqnarray}
\Sigma_{\mathrm{L}}(i\varepsilon_n,\mathbf{k}) &=& -
e^2T\sum_{i\omega_{m}}\frac{1}{2}\mathrm{Tr}
\Big[\mathbf{1}\cdot\int\frac{d^2\mathbf{q}}{(2\pi)^{2}}\nonumber
\\
&&\times G_{0}(i\varepsilon_{n}+i\omega_{m},\mathbf{k+q})
D_{00}(i\omega_{m},\mathbf{q})\Big],\label{eqn:SelfEnergyL}
\\
\Sigma_{\mathrm{T}}(i\varepsilon_n,\mathbf{k}) &=&
e^2T\sum_{i\omega_{m}}\frac{1}{2}\mathrm{Tr}
\Big[\mathbf{1}\cdot\int\frac{d^2\mathbf{q}}{(2\pi)^{2}}\nonumber
\\
&& \times \sigma_{i}
G_{0}(i\varepsilon_{n}+i\omega_{m},\mathbf{k+q})\sigma_{j}
D_{ij}(i\omega_{m},\mathbf{q})\Big], \nonumber \\ \label{eqn:SelfEnergyT}
\end{eqnarray}
are the contributions from the longitudinal and transverse
components of the gauge field, respectively. The damping rate of
massless Dirac fermion will be obtained by making analytic
continuation, $i\varepsilon_n\rightarrow\varepsilon+i\delta$, as
\begin{equation}
\Sigma(\varepsilon,\mathbf{k}) = \Sigma_{\mathrm{L}}(\varepsilon,\mathbf{k}) +
\Sigma_{\mathrm{T}}(\varepsilon,\mathbf{k}),
\end{equation}
and then taking the imaginary part,
$\mathrm{Im}\Sigma(\varepsilon,\mathbf{k})$.

When the Fermi level lies exactly at the Dirac point, the states
below the point are all occupied while the states beyond it are all
empty (see Fig.1(a)). In this state, the chemical potential is
usually defined as zero, $\mu = 0$. In a previous paper, we studied
the Dirac fermion damping rate and found that it behaves as
$\mathrm{Im}\Sigma(\varepsilon) \propto \varepsilon^{1/2}$ at zero
temperature, which is a typical non-Fermi liquid behavior. Once the
fermion density increases starting from the Dirac point, the system
develops a finite chemical potential $\mu$ (Fig.1(b)). Now the
system has a finite but small Fermi surface. The density of states
of fermions at the Fermi level has a finite quantity. Therefore, at
finite $\mu$, the gauge field may lead to very different behaviors
for the fermion damping rate.

In order to know how fermion damping rate varies with $\mu$, we will
explicitly calculate the fermion self-energy. To this end, we first
calculate the polarization functions
$\Pi_{00}(i\omega_{m},\mathbf{q})$ and
$\Pi_{ii}(i\omega_{m},\mathbf{q})$.

\section{Computation of polarization functions}

The polarization functions contributed by the massless Dirac
fermions deserve careful exploration since they determine or are
directly related to many important physical quantities. For
instance, the dynamical screening effect of collective particle-hole
excitations on the gauge or Coulomb interaction between Dirac
fermions can only be studied by the polarization functions.
Physically, such effect describes the damping of gauge boson in the
many-body background composed of massless Dirac fermions. In
addition, according to the Kubo formula in transport theory, various
conductivities are all given by their corresponding current-current
correlation functions, which in form are analogous to the
polarization functions.

In this section, we briefly outline the computational steps and
present the complete expressions for polarization functions
$\Pi_{00}$ and $\Pi_{\bot}$ in the presence of finite chemical
potential at both zero and finite temperature. The analytical
expressions will be useful to any work that relies on the properties
of polarization functions of two-dimensional Dirac fermions.

\subsection{Temporal component $\Pi_{00}(\omega,\mathbf{q},T)$
\label{Sec:Pi00Im}}

To calculate the temporal component of polarization function
$\Pi_{00}(i\omega_{m},\mathbf{q})$, we first introduce the spectral
representation
\begin{equation}
G_0\left(i\varepsilon_{n},\mathbf{k}\right) =
-\int_{-\infty}^{+\infty}\frac{d\omega_1}{\pi}
\frac{\mathrm{Im}\left[G_0(\omega_1,\mathbf{k})\right]}{i\varepsilon_{n}
- \omega_1}\label{eqn:frequencyR}
\end{equation}
and then sum over the frequency, which yields

\begin{widetext}

\begin{eqnarray}
\Pi_{00}\left(i\omega_{m},\mathbf{q}\right) &=& -Ne^2
\int\frac{d^2\mathbf{k}}{(2\pi)^2}
\mathrm{Tr}\left[\int_{-\infty}^{+\infty}
\frac{d\omega_1}{\pi}\mathrm{Im}\left[G_0(\omega_1,\mathbf{k})\right]
\int_{-\infty}^{+\infty} \frac{d\omega_2}{\pi}
\mathrm{Im}\left[G_0(\omega_2,\mathbf{k+q})\right]\right] \nonumber
\\
&&\times \frac{n_F(\omega_1) -
n_F(\omega_2)}{\omega_1-\omega_2+i\omega_{m}}.
\label{eqn:Polarization00BA}
\end{eqnarray}
It is convenient to make the analytic continuation $i\omega_{m}
\rightarrow \omega+i\delta$ at this stage:
\begin{equation}
\frac{1}{\omega_1-\omega_2+i\omega_n}\rightarrow
\frac{1}{\omega_1-\omega_2+\omega+i\delta}
=P\frac{1}{\omega_1-\omega_2+\omega} -
i\pi\delta(\omega_1-\omega_2+\omega).
\label{eqn:AnalyticContinuation}
\end{equation}
The imaginary part of the retarded polarization function is
\begin{eqnarray}
\mathrm{Im}\Pi_{00}(\omega,\mathbf{q}) &=& N\pi
e^2\int\frac{d^2\mathbf{k}}{(2\pi)^2}
\mathrm{Tr}\left[\int_{-\infty}^{+\infty}
\frac{d\omega_1}{\pi}\mathrm{Im}\left[G_0(\omega_1,\mathbf{k})\right]
\int_{-\infty}^{+\infty}
\frac{d\omega_2}{\pi}\mathrm{Im}\left[G_0\left(
\omega_2,\mathbf{k+q}\right)\right]\right] \nonumber \\
&&\times \left[n_F(\omega_1) - n_F(\omega_2)\right]
\delta\left(\omega_1-\omega_2+\omega\right).
\label{eqn:polarization00Im}
\end{eqnarray}
Here the imaginary part of retarded fermion Green function is given
by
\begin{eqnarray}
\mathrm{Im}\left[G_0(\omega,\mathbf{k})\right] &=& \mathrm{Im}
\left[\frac{1}{\omega+\mu-\mathbf{\sigma}\cdot\mathbf{k}+i\delta}\right]
\nonumber \\
&=&\mathrm{Im}\left[\frac{\omega+\mu+\mathbf{\sigma}\cdot\mathbf{k}}
{\left(\omega+\mu\right)^2-\left|\mathbf{k}\right|^2
+i\mathrm{sgn}\left(\omega+\mu\right)\delta}\right] \nonumber \\
&=& -\pi\mathrm{sgn}(\omega+\mu)\left(\omega+\mu +
\mathbf{\sigma\cdot k}\right)\delta\left((\omega+\mu)^2 -
|\mathbf{k}|^2\right) \nonumber\\
&=& -\pi\mathrm{sgn}(\omega+\mu)\left(\omega+\mu +
\mathbf{\sigma\cdot k}\right)\frac{1}{2
|\mathbf{k}|}\left[\delta\left(\omega+\mu+|\mathbf{k}|\right)
+\delta\left(\omega+\mu-|\mathbf{k}|\right)\right].
\label{eqn:progagatorIm}
\end{eqnarray}
After tedious computation, we finally have
\begin{equation}
\mathrm{Im}\Pi_{00}\left(\omega,\mathbf{q},T\right)
\\
= \left\{
\begin{array}{ll}
 \sum_{\alpha=\pm1}\mathrm{sgn}(\omega)
\frac{Ne^2}{8\pi }\frac{|\mathbf{q}|^2}
{\sqrt{\omega^2-|\mathbf{q}|^2} }\int_{-1} ^{1} dx\sqrt{1- x^2}
\left[\delta_{\alpha,1}-\frac{1}{1+e^{\frac{|\mathbf{q}|x +
|\omega|-2\alpha\mu}{2T}}}\right]
\\
\hspace{9cm}\mbox{when}\qquad|\omega| > |\mathbf{q}|,
   \\
   \\
\sum_{\alpha=\pm1}\mathrm{sgn}(\omega)
\frac{Ne^2}{8\pi}\frac{|\mathbf{q}|^2}
{\sqrt{|\mathbf{q}|^2-\omega^2}}\int_{1}^{+\infty}dx
\left[\frac{\sqrt{x^2-1}}{1+e^{\frac{|\mathbf{q}|x-|\omega|-2\alpha\mu}{2T}}}
-\frac{\sqrt{x^2-1}}{1+e^{\frac{|\mathbf{q}|x+|\omega|-2\alpha\mu}{2T}}}\right]
\\
\hspace{9cm}\mbox{when}\quad|\omega| < |\mathbf{q}|.
\end{array}
\right.\label{eqn:Polarization00ImFinitT}
\end{equation}

We now calculate the real part of temporal component of polarization
function. The whole computation is much more complicated than the
imaginary part. From Eq.(\ref{eqn:Polarization00BA}) and
Eq.(\ref{eqn:AnalyticContinuation}), we have
\begin{eqnarray}
\mathrm{Re}\Pi_{00}\left(\omega,\mathbf{q},T\right) &=&
-Ne^2P\int\frac{d^2\mathbf{k}}{(2\pi)^2}\mathrm{Tr}
\left[\int_{-\infty}^{+\infty}\frac{d\omega_1}{\pi}\mathrm{Im}
\left[G_0\left(\omega_1,\mathbf{k}\right)
\right]\int_{-\infty}^{+\infty}\frac{d\omega_2}{\pi}\mathrm{Im}
\left[G_0(\omega_2,\mathbf{k+q})\right]\right] \nonumber
\\
&& \times \frac{n_F(\omega_1) - n_F(\omega_2)}
{\omega_1-\omega_2+\omega} \nonumber \\
&=& -Ne^2\int\frac{d^2\mathbf{k}}{(2\pi)^2}\mathrm{Tr}
\left[\int_{-\infty}^{+\infty}\frac{d\omega_1}{\pi}\mathrm{Im}
\left[G_0\left(\omega_1,\mathbf{k}\right)\right] n_F(\omega_1)
P\int_{-\infty}^{+\infty}\frac{d\omega_2}{\pi}
\frac{\mathrm{Im}\left[G_0(\omega_2,\mathbf{k+q})\right]}
{\omega_1+\omega-\omega_2}\right] \nonumber \\
&& -Ne^2\int\frac{d^2\mathbf{k}}{(2\pi)^2}\mathrm{Tr}
\left[\int_{-\infty}^{+\infty}\frac{d\omega_2}{\pi}\mathrm{Im}
\left[G_0(\omega_2,\mathbf{k+q})\right] n_F(\omega_2)
P\int_{-\infty}^{+\infty}\frac{d\omega_1}{\pi}
\frac{\mathrm{Im}\left[G_0(\omega_1,\mathbf{k})\right]}
{\omega_2-\omega-\omega_1}\right].
\end{eqnarray}
Using the Krames-Kronig relation
\begin{equation}
\mathrm{Re}\left[G_0(\omega,\mathbf{k})\right] =
-P\int_{-\infty}^{+\infty} \frac{d\omega'}{\pi}
\frac{\mathrm{Im}\left[G_0
(\omega',\mathbf{k})\right]}{\omega-\omega'},\label{eqn:KKrelation}
\end{equation}
the above expression can now be converted to
\begin{eqnarray}
\mathrm{Re}\Pi_{00}(\omega,\mathbf{q},T) &=&
Ne^2\int\frac{d^2\mathbf{k}}{(2\pi)^2}\int_{-\infty}^{+\infty}
\frac{d\omega_1}{\pi}n_F(\omega_1)\mathrm{Tr}
\left[\mathrm{Im}\left[G_0(\omega_1,\mathbf{k})
\right]\mathrm{Re}\left[G_0(\omega_1+\omega,\mathbf{k+q})
\right]\right] \nonumber \\
&& +Ne^2\int\frac{d^2\mathbf{k}}{(2\pi)^2}\int_{-\infty}^{+\infty}
\frac{d\omega_1}{\pi}n_F(\omega_1)\mathrm{Tr}
\left[\mathrm{Im}\left[G_0(\omega_1,\mathbf{k+q})\right]
\mathrm{Re}\left[G_0(\omega_1-\omega,\mathbf{k}) \right]\right],
\label{eqn:Polarization00Re}
\end{eqnarray}
where
\begin{equation}
\mathrm{Re}G_0\left(\omega,\mathbf{k}\right) =
\left(\omega+\mu+\mathbf{\sigma\cdot k}\right)
P\frac{1}{\left(\omega+\mu\right)^2 - \left|\mathbf{k}\right|^2}.
\label{eqn:PropagatorRe}
\end{equation}
From this equation, we obtain the following expression:
\begin{equation}
\mathrm{Re}\Pi_{00}\left(\omega,\mathbf{q},T\right)
=-\frac{Ne^2}{2\pi}\int_0^{+\infty}d|\mathbf{k}|+\left\{
\begin{array}{ll}
\sum_{\alpha=\pm1}\left\{\frac{Ne^2T\ln\left(1 +
e^{\frac{\alpha\mu}{T}}\right)}{2\pi}
-\frac{Ne^2}{8\pi}\frac{|\mathbf{q}|^2}{\sqrt{\omega^2 -
|\mathbf{q}|^2}}\int_{1}^{+\infty}dx\sqrt{x^2-1}\right.
\\
\left.\times\left[\frac{1}{1+e^{\frac{\left||\mathbf{q}|x -
|\omega|\right| - 2\alpha\mu}{2T}}}
-\frac{1}{1+e^{\frac{|\mathbf{q}|x + |\omega| -
2\alpha\mu}{2T}}}\right]\right\}
\\
\hspace{5cm}\mbox{when}\quad |\omega| > |\mathbf{q}|,
\\
\\
\sum_{\alpha\pm1}\left\{\frac{Ne^2T\ln\left(1+e^{\frac{\alpha\mu}{T}}\right)}{2\pi}
+\frac{Ne^2}{8\pi}\frac{|\mathbf{q}|^2}{\sqrt{|\mathbf{q}|^2-\omega^2}}
\int_{-1}^{1}dx\sqrt{1-x^2}\right.
\\
\left.\times\left[\delta_{\alpha,1} -\frac{1}{1+e^{\frac{\left|
|\mathbf{q}|x + |\omega|\right|-2\alpha\mu}{2T}}}\right]\right\}
\\
\hspace{5cm}\mbox{when}\quad |\omega| < |\mathbf{q}|.
\end{array}
\right.\label{eqn:Polarization00ReFinitT}
\end{equation}
Notice there appears a divergent term
\begin{equation}
I_{\mathrm{Singular}}=-\frac{Ne^2}{2\pi}\int_0^{+\infty}d|\mathbf{k}|.
\end{equation}
To remove this divergence, here we employ the regularization scheme
that was proposed in \cite{Franz02}. This scheme states that the
gauge field must remain massless so that it should satisfy
\begin{equation}
\Pi_{\mu\nu}\left(\omega\rightarrow0,|\mathbf{q}|\rightarrow0,\mu=0,T=0
\right) = 0.
\end{equation}
Now the polarization function can be re-defined as
\begin{equation}
\Pi_{\mu\nu}(\omega,\mathbf{q},T) -
\Pi_{\mu\nu}\left(\omega\rightarrow0,|\mathbf{q}|\rightarrow0,\mu=0,T=0\right).
\end{equation}
After this regularization, the singular term can be simply dropped
from $\mathrm{Re}\Pi_{00}\left(\omega,\mathbf{q},T\right)$.

\subsection{Transverse component $\Pi_{\bot}\left(\omega,\mathbf{q},T\right)$}

Proceeding as we have done in the above, we found that the imaginary
and real parts of $\Pi_{ii}\left(\omega,\mathbf{q},T\right)$ have
the expressions:
\begin{equation}
\mathrm{Im}\Pi_{ii}\left(\omega,\mathbf{q},T\right) = \left\{
\begin{array}{ll}
\sum_{\alpha=\pm1}\left\{-\mathrm{sgn}(\omega)
\frac{Ne^2}{8\pi}\frac{|\mathbf{q}|^2}{\sqrt{\omega^2 -
|\mathbf{q}|^2}} \int_{-1}^{1} dx\sqrt{1-x^2}
\left[\delta_{\alpha,1}-\frac{1}{1+e^{\frac{|\mathbf{q}|x+|\omega|-2\alpha\mu}{2T}}}\right]
\right.
\\
\left.-\mathrm{sgn}(\omega)\frac{Ne^2}{8\pi}\sqrt{\omega^2-|\mathbf{q}|^2}
\int_{-1}^{1}dx\frac{1}{\sqrt{1-x^2}}
\left[\delta_{\alpha,1}-\frac{1}{1+e^{\frac{|\mathbf{q}|x+|\omega|-2\alpha\mu}{2T}}}\right]
\right\}
\\
\hspace{9cm}\mbox{when}\qquad|\omega|>|\mathbf{q}|,
   \\
   \\
\sum_{\alpha=\pm1}\left\{-\mathrm{sgn}(\omega)
\frac{Ne^2}{8\pi}\frac{|\mathbf{q}|^2}{\sqrt{|\mathbf{q}|^2-\omega^2}}
\int_{1}^{+\infty}
dx\left[\frac{\sqrt{x^2-1}}{1+e^{\frac{|\mathbf{q}|x-|\omega|-2\alpha\mu}{2T}}}
-\frac{\sqrt{x^2-1}}{1+e^{\frac{|\mathbf{q}|x+|\omega|-2\alpha\mu}{2T}}}\right]
\right. \nonumber
\\
\left.-\mathrm{sgn}(\omega)\frac{Ne^2}{8\pi}\sqrt{|\mathbf{q}|^2-\omega^2
}\int_{1}^{+\infty}
dx\frac{1}{\sqrt{x^2-1}}\left[\frac{1}{1+e^{\frac{|\mathbf{q}|x-|\omega|-2\alpha\mu}{2T}}}
-\frac{1}{1+e^{\frac{|\mathbf{q}|x+|\omega|-2\alpha\mu}{2T}}}\right]
\right\}
\\
\hspace{9cm}\mbox{when}\qquad |\omega| < |\mathbf{q}|.
\end{array}
\right.\label{eqn:PolarizationiiImFinitT}
\end{equation}

\begin{equation}
\mathrm{Re}\Pi_{ii}\left(\omega,\mathbf{q},T\right)
=\left\{
\begin{array}{ll}
\sum_{\alpha=\pm1}\left\{\frac{Ne^2}{8\pi}
\frac{|\mathbf{q}|^2}{\sqrt{\omega^2-|\mathbf{q}|^2}}
\int_{1}^{+\infty}dx
\left[\frac{\sqrt{x^2-1}}{1+e^{\frac{\left||\mathbf{q}|x-|\omega|\right|
- 2\alpha\mu}{2T}}}
-\frac{\sqrt{x^2-1}}{1+e^{\frac{|\mathbf{q}|x+|\omega| -
2\alpha\mu}{2T}}}\right]\right.
\\
\\
\left.-\frac{Ne^2}{8\pi}\sqrt{\omega^2 -
|\mathbf{q}|^2}\int_{1}^{+\infty}dx\frac{1}{\sqrt{x^2-1}}
\left[\frac{1}{1+e^{\frac{\left||\mathbf{q}|x-|\omega|\right|-2\alpha\mu}{2T}}}
-\frac{1}{1+e^{\frac{|\mathbf{q}|x+|\omega|-2\alpha\mu}{2T}}}\right]
\right\}
\\
\hspace{9cm}\mbox{when}\quad |\omega| > |\mathbf{q}|,
\\
\\
\sum_{\alpha=\pm1}\left\{-\frac{Ne^2}{8\pi}
\frac{|\mathbf{q}|^2}{\sqrt{|\mathbf{q}|^2-\omega^2}}
\int_{-1}^{1}dx\sqrt{1-x^2}
\left[\delta_{\alpha,1}-\frac{1}{1+e^{\frac{\left||\mathbf{q}|x+|\omega|\right|
- 2\alpha\mu}{2T}}}\right]\right.
\\
\\
\left.+\frac{Ne^2}{8\pi}\sqrt{|\mathbf{q}|^2-\omega^2}
\int_{-1}^{1}dx\frac{1}{\sqrt{1-x^2}}
\left[\delta_{\alpha,1}-\frac{1}{1+e^{\frac{\left||\mathbf{q}|x+|\omega|\right|
- 2\alpha\mu}{2T}}}\right]\right\}
\\
\hspace{9cm}\mbox{when}\quad |\omega| < |\mathbf{q}|.
\end{array}
\right.\label{eqn:PolarizationiiReFinitT}
\end{equation}

According to Eq.(\ref{eqn:PolarizationBot}), the retarded transverse polarization function
is decomposed as
\begin{equation}
\Pi_{\bot}(\omega,\mathbf{q},T) = \Pi_{ii}(\omega,\mathbf{q},T) +
\frac{\omega^2}{|\mathbf{q}|^2} \Pi_{00}(\omega,\mathbf{q},T),
\end{equation}
which can be written more explicitly as
\begin{eqnarray}
\mathrm{Im}\Pi_{\bot}(\omega,\mathbf{q},T) &=&
\mathrm{Im}\Pi_{ii}(\omega,\mathbf{q},T) + \frac{\omega^2}{
|\mathbf{q}|^2} \mathrm{Im}\Pi_{00}(\omega,\mathbf{q},T),
\label{eqn:PolarizationBotIm}
\\
\mathrm{Re}\Pi_{\bot}(\omega,\mathbf{q},T) &=&
\mathrm{Re}\Pi_{ii}(\omega,\mathbf{q},T) + \frac{\omega^2}{
|\mathbf{q}|^2}
\mathrm{Re}\Pi_{00}(\omega,\mathbf{q},T).\label{eqn:PolarizationBotRe}
\end{eqnarray}
Using the results presented above, it is easy to get that
\begin{equation}
\mathrm{Im}\Pi_{\bot}(\omega,\mathbf{q},T) = \left\{
\begin{array}{ll}
-\sum_{\alpha=\pm1}\mathrm{sgn}(\omega)
\frac{Ne^2}{8\pi}\sqrt{\omega^2 - |\mathbf{q}|^2}
\int_{-1}^{1} dx\frac{x^2}{\sqrt{1-x^2}}
\left[\delta_{\alpha,1}-\frac{1}{1+e^{\frac{|\mathbf{q}|x +
|\omega|-2\alpha\mu}{2T}}}\right]
\\
\hspace{9cm}\mbox{when}\qquad |\omega| > |\mathbf{q}|,
   \\
   \\
-\sum_{\alpha=\pm1}\mathrm{sgn}(\omega)\frac{Ne^2}{8\pi}
\sqrt{|\mathbf{q}|^2-\omega^2 }\int_{1}^{+\infty}
dx\frac{x^2}{\sqrt{x^2-1}}
\left[\frac{1}{1+e^{\frac{|\mathbf{q}|x-|\omega|-2\alpha\mu}{2T}}}\right.
\\
\left.-\frac{1}{1+e^{\frac{|\mathbf{q}|x+|\omega|-2\alpha\mu}{2T}}}\right]
\\
\hspace{9cm}\mbox{when}\qquad |\omega| < |\mathbf{q}|.
\end{array}
\right.\label{eqn:PolarizationBotImFinitT}
\end{equation}
\begin{equation}
\mathrm{Re}\Pi_{\bot}(\omega,\mathbf{q},T) = \left\{
\begin{array}{ll}
\sum_{\alpha=\pm1}\left\{\frac{Ne^2T\ln\left(1+e^{\frac{\alpha\mu}{T}}\right)}{2\pi}
\frac{\omega^2}{|\mathbf{q}|^2} -
\frac{Ne^2}{8\pi}\sqrt{\omega^2-|\mathbf{q}|^2}
\int_{1}^{+\infty}dx\frac{x^2}{\sqrt{x^2-1}}\right.
\\
\left.\times\left[\frac{1}{1+e^{\frac{\left||\mathbf{q}|x-|\omega|\right|-2\alpha\mu}{2T}}}
-\frac{1}{1+e^{\frac{|\mathbf{q}|x+|\omega|-2\alpha\mu}{2T}}}\right]\right\}
\\
\hspace{9cm}\mbox{when}\qquad |\omega| > |\mathbf{q}|,
   \\
   \\
\sum_{\alpha=\pm1}\left\{\frac{Ne^2T\ln\left(1+e^{\frac{\alpha\mu}{T}}\right)}{2\pi}
\frac{\omega^2}{|\mathbf{q}|^2} +
\frac{Ne^2}{8\pi}\sqrt{|\mathbf{q}|^2-\omega^2}
\int_{-1}^{1}dx\frac{x^2}{\sqrt{1-x^2}}\right.
\\
\left.\times\left[\delta_{\alpha,1}-\frac{1}{1+e^{\frac{\left||\mathbf{q}|x
+ |\omega|\right|-2\alpha\mu}{2T}}}\right]\right\}
\\
\hspace{9cm}\mbox{when}\qquad |\omega| < |\mathbf{q}|.
\end{array}
\right.\label{eqn:PolarizationBotReFinitT}
\end{equation}

\section{Fermion damping rate at zero temperature}

In this section, we calculate the fermion damping rate at zero
temperature. We first consider the transverse contribution of gauge
field to fermion damping rate. To do this, we will substitute the
transverse gauge propagator Eq.(\ref{eqn:Photonij}) to transvrese
self-energy function Eq.(\ref{eqn:SelfEnergyT}). Here, it is
convenient to introduce the following spectral representations:
\begin{eqnarray}
G_0\left(i\varepsilon_{n}+i\omega_{m},\mathbf{k+q}\right) &=&
-P\int_{-\infty}^{+\infty}\frac{d\omega_1}{\pi}\frac{\mathrm{Im}
\left[G_0\left(\omega_1,\mathbf{k+q}\right)\right]}
{i\varepsilon_{n}+i\omega_{m}-\omega_1}, \\
\frac{1}{|\mathbf{q}|^2+\omega_{m}^{2}
+\Pi_{\bot}\left(i\omega_{m},|\mathbf{q}|\right)} &=&
-P\int_{-\infty}^{+\infty}\frac{d\omega_2}{\pi}
\frac{1}{i\omega_{m}-\omega_2}
\mathrm{Im}\left[\frac{1}{|\mathbf{q}|^2-\omega_2^2 -
i\mathrm{sgn}\left(\omega_2\right)\delta +
\Pi_{\bot}\left(\omega_2,|\mathbf{q}|\right)}\right].
\end{eqnarray}
After carrying out the summation over $\omega_m$, we can get
\begin{eqnarray}
\Sigma_{\mathrm{T}}\left(i\varepsilon_{n},\mathbf{k}\right) &=&
-e^2\frac{1}{2} \mathrm{Tr}\Bigg[\mathbf{1}\cdot
\int\frac{d^2\mathbf{q}}{\left(2\pi\right)^2}\sigma_i
\int_{-\infty}^{+\infty}\frac{d\omega_1}{\pi}\mathrm{Im}\left[G_0
\left(\omega_1,\mathbf{k+q}\right)\right]\sigma_j \left(\delta_{ij}
- q_iq_j/|\mathbf{q}|^2\right)\int_{-\infty}^{+\infty}
\frac{d\omega_2}{\pi} \nonumber \\
&& \times\mathrm{Im}\left[\frac{1}{|\mathbf{q}|^2-\omega_2^2
-i\mathrm{sgn}\left(\omega_2\right)\delta
+\Pi_{\bot}\left(\omega_2,|\mathbf{q}|\right)}\right]
\frac{n_B\left(\omega_2\right)+n_F\left(\omega_1\right)}
{i\varepsilon_{n}+\omega_2-\omega_1}\Bigg].
\end{eqnarray}
After analytic continuation
$i\varepsilon_{n}\rightarrow\varepsilon+i\delta$, we have
\begin{equation}
\frac{1}{i\varepsilon_{n}+\omega_2-\omega_1}
\rightarrow\frac{1}{\varepsilon+\omega_2-\omega_1+i\delta} =
P\frac{1}{\varepsilon+\omega_2-\omega_1} -
i\pi\delta\left(\varepsilon+\omega_2-\omega_1\right),
\end{equation}
and the imaginary part of fermion self-energy becomes
\begin{eqnarray}
\mathrm{Im}\Sigma_{\mathrm{T}}(\varepsilon,\mathbf{k},T) &=&
-e^2\int\frac{d^2\mathbf{q}}{(2\pi)^2}\mathrm{Im}\left[\frac{1}
{|\mathbf{q}|^2-(|\mathbf{k+q}|-\mu-\varepsilon)^2
-i\delta\mathrm{sgn}(|\mathbf{k+q}|-\mu-\varepsilon)+\Pi_{\bot}
(|\mathbf{k+q}|-\mu-\varepsilon,|\mathbf{q}|)}\right] \nonumber \\
&&\times\left[n_B(|\mathbf{k+q}|-\mu-\varepsilon)+n_F(|\mathbf{k+q}|-\mu)\right],
\end{eqnarray}
where Eq.(\ref{eqn:progagatorIm})
was used. We now introduce a new variable $\mathbf{k'} =
\mathbf{k}+\mathbf{q}$, and then have
\begin{eqnarray}
\mathrm{Im}\Sigma_{\mathrm{T}}(\varepsilon,\mathbf{k},T) &=&
-\frac{e^2}{4\pi^2}\int_{0}^{+\infty}d|\mathbf{k'}||\mathbf{k'}|
\int_{0}^{2\pi}d\theta \nonumber \\
&&\times \mathrm{Im}\left[\frac{1}
{|\mathbf{k'-k}|^2-(|\mathbf{k'}|-\mu-\varepsilon)^2
-i\delta\mathrm{sgn}(|\mathbf{k'}|-\mu-\varepsilon)+\Pi_{\bot}
(|\mathbf{k'}|-\mu-\varepsilon,|\mathbf{k'-k}|)}\right]\nonumber \\
&& \times \left[n_B(|\mathbf{k'}|-\mu-\varepsilon) +
n_F(|\mathbf{k'}|-\mu)\right],
\end{eqnarray}
where $\theta$ is the angle between $\mathbf{k}$ and $\mathbf{k'}$.
Without lose of generality, we suppose that $\varepsilon > 0$.

Now we focus on the zero temperature limit and consider the case at
finite temperature in the next section. At $T=0$, the contribution
function reduces to
\begin{eqnarray}
n_B\left(|\mathbf{k'}|-\mu-\varepsilon\right) +
n_F\left(|\mathbf{k'}|-\mu\right) =
-\theta\left(|\mathbf{k'}|-\mu\right)
\theta\left(\mu+\varepsilon-|\mathbf{k'}|\right),
\end{eqnarray}
so the transverse damping rate reduces to
\begin{eqnarray}
\mathrm{Im}\Sigma_{\mathrm{T}}(\varepsilon,\mathbf{k}) &=&
\frac{e^2}{4\pi^2}\int_{\mu}^{\mu+\varepsilon}d|\mathbf{k'}||\mathbf{k'}|
\int_{0}^{2\pi}d\theta\mathrm{Im}\left[\frac{1}
{|\mathbf{k'-k}|^2-(|\mathbf{k'}|-\mu-\varepsilon)^2
+i\delta+\Pi_{\bot}
(|\mathbf{k'}|-\mu-\varepsilon,|\mathbf{k'-k}|)}\right].
\end{eqnarray}
In general, there are two kinds of approximations: on-shell
approximation and fixed-momentum approximation. We now consider the
on-shell approximation
\begin{eqnarray}
\varepsilon = \xi_{\mathbf{k}} = \varepsilon_{\mathbf{k}}-\mu =
|\mathbf{k}|-\mu,
\end{eqnarray}
and convert the damping rate to
\begin{eqnarray}
\mathrm{Im}\Sigma_{\mathrm{T}}(\xi_{\mathbf{k}})
=\frac{e^2}{4\pi^2}\int_{\mu}^{\mu+\xi_{\mathbf{k}}}
d|\mathbf{k'}||\mathbf{k'}|
\int_{0}^{2\pi}d\theta\mathrm{Im}\left[\frac{1}
{|\mathbf{k'-k}|^2-(|\mathbf{k'}|-|\mathbf{k}|)^2
+i\delta+\Pi_{\bot}(|\mathbf{k'}|-|\mathbf{k}|,|\mathbf{k'-k}|)}\right].
\label{eqn:ZeroTransverse}
\end{eqnarray}
The fixed momentum approximation will be discussed later.

\end{widetext}

To proceed, we will substitute the analytical expression of
polarization function $\Pi_{\bot}(\omega,|\mathbf{q}|)$ obtained in
the last section into this formula. At the $T = 0$ limit, the
integration over parameter $x$ in Eq.(\ref{eqn:PolarizationBotImFinitT})
and Eq.(\ref{eqn:PolarizationBotReFinitT}) can be analytically carried out. The
expression for $\Pi_{\bot}(\omega,|\mathbf{q}|)$ at $T=0$ is
presented in the Appendix. Such expression is clearly too
complicated to be used. In order to get analytical results for
fermion damping rate, it is necessary to make proper approximations
to $\Pi_{\bot}(\omega,|\mathbf{q}|)$.

In the present problem, it is important to observe that the dominant
contribution of the above integral comes form the region
$|\omega|\ll |\mathbf{q}|$ and $|\mathbf{q}|\ll\mu$ in
$\Pi_{\bot}(\omega,|\mathbf{q}|)$, so that we can simply the
polarization functions by restricting the energy-momentum to this
region. This approximation method was used by many authors
previously \cite{Holstein, Lee89, Blok93, Polchinski, Metzner}. In this
region, the polarization function can be significantly simplified
and is given by
\begin{eqnarray}
&& \mathrm{Re}\Pi_{\bot}(\omega,|\mathbf{q}|) =
\frac{Ne^2\mu}{2\pi}\frac{\omega^2}{|\mathbf{q}|^2},\label{eqn:PolarizationBotReSmallESmallQ}
\\
&& \mathrm{Im}\Pi_{\bot}(\omega,|\mathbf{q}|) \approx
-\mathrm{sgn}(\omega)\frac{Ne^2\mu}{2\pi}
\frac{|\omega|}{|\mathbf{q}|}.\label{eqn:PolarizationBotImSmallESmallQ}
\end{eqnarray}
If we take the static limit, $\omega \rightarrow 0$, both the real
and imaginary parts of the transverse polarization function
vanishes, $\Pi_{\bot}(\omega,|\mathbf{q}|)\rightarrow 0$. This
implies that the transverse gauge field remains massless even after
including the dynamical screening effect due to particle-hole
excitations. This property is robust against higher order
corrections and indeed a consequence of gauge invariance. However,
the chemical potential $\mu$ does affect the transverse gauge
interaction between Dirac fermions and thus should affect the
fermion damping rate. Substituting the above expressions for
$\Pi_{\bot}(\omega,|\mathbf{q}|)$ into
Eq.(\ref{eqn:ZeroTransverse}), we finally get
\begin{eqnarray}
\mathrm{Im}\Sigma_{\mathrm{T}}(\xi_{\mathbf{k}}) \approx
C\left(\mu\right)\xi_{\mathbf{k}}^{\frac{2}{3}},
\end{eqnarray}
where
\begin{equation}
C(\mu) = -\frac{\sqrt[3]{2}
e^{\frac{4}{3}}}{8\sqrt{3}\pi^{\frac{2}{3}}
N^{\frac{1}{3}}\mu^{\frac{1}{3}}}.
\end{equation}
Apparently, this damping rate displays non-Fermi liquid behavior at
zero temperature.

We now consider the longitudinal contribution to fermion damping
rate. Starting from Eq.(\ref{eqn:Photon00}) and
Eq.(\ref{eqn:SelfEnergyL}) and then using the same steps presented
in the above, we write the longitudinal damping rate as
\begin{eqnarray}
\mathrm{Im}\Sigma_{\mathrm{L}}\left(\xi_{\mathbf{k}}\right) &=&
-\frac{e^2}{4\pi^2}\int_{\mu}^{\mu+\xi_{\mathbf{k}}}d|\mathbf{k'}|
|\mathbf{k'}|\int_0^{2\pi}d\theta \nonumber \\
&& \times \mathrm{Im} \left[\frac{1}{|\mathbf{k'-k}|^2 +
\Pi_{00}\left(|\mathbf{k'}|-|\mathbf{k}|,|\mathbf{k'-k}|\right)}\right].
\nonumber \\
\label{eqn:ZeroLognitudinal}
\end{eqnarray}
As $\Pi_{\bot}(\omega,|\mathbf{q}|)$, the polarization function
$\Pi_{00}(\omega,|\mathbf{q}|)$ is also too complicated even at $T =
0$ (see Appendix). In the region $|\omega|\ll |\mathbf{q}|$ and
$|\mathbf{q}|\ll \mu$, we have the following simplified expressions
for $\Pi_{00}(\omega,|\mathbf{q}|)$:
\begin{eqnarray}
&& \mathrm{Re}\Pi_{00}(\omega,|\mathbf{q}|) = \frac{Ne^2\mu}{2\pi},
\label{eqn:Polarization00ReSmallESmallQ}
\\
&&\mathrm{Im}\Pi_{00}(\omega,|\mathbf{q}|)
\approx\mathrm{sgn}(\omega)\frac{Ne^2\mu}{2\pi}\frac{|\omega|}{|\mathbf{q}|}.
\label{eqn:Polarization00ImSmallESmallQ}
\end{eqnarray}
In the static limit, $\omega \rightarrow 0$, the imaginary part
$\mathrm{Im}\Pi_{00}(\omega,|\mathbf{q}|)$ vanishes but the real
part $\mathrm{Re}\Pi_{00}(\omega,|\mathbf{q}|)$ is a constant.
Therefore, from Eq.(\ref{eqn:Photon00}), the temporal component of
gauge field propagator is found to be
\begin{eqnarray}
D_{00}(\omega = 0,\mathbf{q}) = \frac{1}{|\mathbf{q}|^2 +
\frac{Ne^2\mu}{2\pi}}
\end{eqnarray}
in the static limit. Comparing with the transverse component of
gauge field propagator defined by Eq.(\ref{eqn:Photonij}),
Eq.(\ref{eqn:PolarizationBotReSmallESmallQ}) and
Eq.(\ref{eqn:PolarizationBotImSmallESmallQ}), the temporal component
has a static screening and chemical potential $\mu$ defines the
Debye screening length. It reflects the effect of particle-hole
excitations on the initially long-range temporal gauge interaction.
This is the key difference between Dirac fermion systems with zero
and finite chemical potential. The short-range temporal gauge
interaction is expected to produce only normal Fermi liquid
behavior. Substituting them into Eq.(\ref{eqn:ZeroLognitudinal}), it
is easy to get
\begin{equation}
\mathrm{Im}\Sigma_{\mathrm{L}}\left(\xi_{\mathbf{k}}\right)
\approx\frac{1}{2\pi N\mu}\xi_{\mathbf{k}}^2
\ln\left(\frac{\xi_{\mathbf{k}}}{\mu}\right).
\end{equation}
This expression vanishes faster than $\xi_{\mathbf{k}}$ as
$\xi_{\mathbf{k}}\rightarrow 0$ and thus is a normal Fermi liquid
behavior. As shown in \cite{WangLiu}, the perturbative result of
zero-temperature fermion damping rate is divergent at $\mu = 0$. The
finite chemical potential eliminates the divergence and at the same
time leads to normal Fermi liquid behavior.

The total fermion damping rate should be
\begin{equation}
\mathrm{Im}\Sigma(\xi_{\mathbf{k}}) =
\mathrm{Im}\Sigma_{\mathrm{T}}(\xi_{\mathbf{k}}) +
\mathrm{Im}\Sigma_{\mathrm{L}}(\xi_{\mathbf{k}}) \approx
C(\mu)\xi_{\mathbf{k}}^{\frac{2}{3}}.
\end{equation}
This result is obtained using the on-shell approximation. We can
alternatively use the fixed momentum approximation. The momentum can
be chosen as the Fermi momentum, so at zero temperature the fermion
damping rate depends only on the energy $\varepsilon$. After
explicit computation, we found that
\begin{eqnarray}
\mathrm{Im}\Sigma_{\mathrm{T}}\left(\varepsilon,|\mathbf{k}|=\mu\right)
&\approx& C(\mu)\varepsilon^{\frac{2}{3}}
\\
\mathrm{Im}\Sigma_{\mathrm{L}}\left(\varepsilon,|\mathbf{k}|=\mu\right)
&\approx&\frac{1}{2\pi
N\mu}\varepsilon^2\ln\left(\frac{\varepsilon}{\mu}\right).
\end{eqnarray}
So the total damping rate is
\begin{equation}
\mathrm{Im}\Sigma\left(\varepsilon,|\mathbf{k}|=\mu\right) \approx
C(\mu)\varepsilon^{\frac{2}{3}}.
\end{equation}
This has the same form as that obtained in the on-shell
approximation, with $\xi_{\mathbf{k}}$ being replaced by
$\varepsilon$.

There are three important features of this damping rate. First, when
we take the $\mu \rightarrow 0$ limit, this result does not reduce
to the $\propto \varepsilon^{1/2}$ result obtained at $\mu = 0$. If
we use the exponent $z$ appearing in the energy-dependence
$\varepsilon^{z}$ of damping rate to characterize the ground state
of the fermion-gauge system, then there is a sudden change of ground
state once $\mu$ departs from zero. It appears that the Dirac
fermion systems exhibit distinct behaviors at zero and finite $\mu$.
This difference arises from the difference in topology of Fermi
surface: at finite $\mu$ the system has a finite one-dimensional
Fermi surface, but at $\mu = 0$ the Fermi surface shrinks to a
zero-dimensional point. Second, at finite $\mu$, as $\mu$ grows from
certain small value, the energy-dependence of fermion damping rate
does not change. Third, at any fixed energy $\varepsilon$ the
fermion damping rate is proportional to $\mu^{-1/3}$, so the Dirac
fermions become more well-defined as chemical potential grows.

The fermion damping rate $\propto \varepsilon^{2/3}$ seems to be a
universal behavior. It has the same energy-dependence as that in
two-dimensional non-relativistic fermion-gauge systems with a large
Fermi surface \cite{Lee89, Blok93, GanWong, Polchinski}. Such
energy-dependence also appears in some two-dimensional electron
systems where fermions interact strongly with fluctuating
ferromagnetic order parameter \cite{Rech} or fluctuating nematic
order \cite{Sunkai, Metlitski}, as well as in two-dimensional
electron systems near a Pomeranchuk instability \cite{Metzner}.

Using the Kramers-Kronig relation, we get the real part of fermion
self-energy
\begin{equation}
\mathrm{Re}\Sigma(\varepsilon) \propto
\sqrt{3}C(\mu)\mathrm{sgn}(\varepsilon)\varepsilon^{\frac{2}{3}}.
\end{equation}
It has the same energy-dependence as the imaginary part. It is easy
to show that the renormalization factor $Z = 0$, which is the
characteristic of a non-Fermi liquid \cite{Giuliani}.

It is also interesting to study the effective QED$_3$ theory without
Maxwell term for the gauge field \cite{Affleck, Kim97}. Now the
propagator for the gauge field in Matsubara formalism has the form
\begin{eqnarray}
D_{00}\left(i\omega_{m},\mathbf{q}\right) &=&
\frac{1}{\Pi_{00}(i\omega_{m},\mathbf{q})},
\\
D_{ij}\left(i\omega_{m},\mathbf{q}\right) &=& \left(\delta_{ij} -
\frac{q_{i}q_{j}}{\mathbf{q}^2}\right)\frac{1}{\Pi_{\bot}
\left(i\omega_{m},\mathbf{q}\right)}.
\end{eqnarray}
Using this propagator, we found that the longitudinal and transverse
fermion damping rates are
\begin{eqnarray}
\mathrm{Im}\Sigma_{\mathrm{T}}(\xi_{\mathbf{k}}) &\approx&
-\frac{\mu}{\pi
N}\int_{0}^{\frac{\xi_{\mathbf{k}}}{\mu}}d\delta'\frac{1}{\delta'},
\\
\mathrm{Im}\Sigma_{\mathrm{L}}\left(\xi_{\mathbf{k}}\right)
&\approx& \frac{1}{2\pi
N\mu}\xi_{\mathbf{k}}^2\ln\left(\frac{\xi_{\mathbf{k}}}{\mu}\right),
\end{eqnarray}
in the on-shell approximation. In the fixed momentum approximation,
we have
\begin{eqnarray}
\mathrm{Im}\Sigma_{\mathrm{T}}(\varepsilon,\left|\mathbf{k}\right|=\mu)
&\approx&-\frac{\mu}{\pi N}\int_{0}^{\frac{\varepsilon}{\mu}}d\delta'\frac{1}{\delta'},
\\
\mathrm{Im}\Sigma_{\mathrm{L}}\left(\varepsilon,\left|\mathbf{k}\right|=\mu\right)
&\approx&\frac{1}{2\pi N\mu}\varepsilon^2\ln\left(\frac{\varepsilon}{\mu}\right).
\end{eqnarray}
Clearly, in both the on-shell and fixed momentum approximations, the
total fermion damping rate is divergent. Note that a similar
divergence also exists at zero chemical potential \cite{WangLiu}. It
seems that such divergences are directly related to the absence of
Maxwell term for the gauge field.

\section{Fermion damping rate at finite temperature}

We now consider the fermion damping rate at finite temperature. The
polarization function at finite $T$ should be used when calculating
the fermion self-energy. Here it will be convenient to adopt an
important approximation. At low temperature $T\ll \mu$, we can still
use the polarization functions obtained at zero temperature. This
approximation was previously employed in Refs. \cite{Reizer89,
Lee89}. In the limit $T \ll \mu$, we can simply choose the upper
boundary value of $|\mathbf{k'}|$ as $\mu + T$. The reason is that
the fermions are primarily scattered into states in the outside of
the Fermi surface, because most of the states on (and below) the
Fermi surface are already occupied by other fermions at low
temperature. The lower limit of $|\mathbf{k'}|$ can be assumed to be
$\mu$. Moreover, at finite temperature, the occupation number
functions can be well simplified as
\begin{eqnarray}
n_B\left(|\mathbf{k'}|-|\mathbf{k}|\right) +
n_F\left(|\mathbf{k'}|-\mu\right) \approx
\frac{T}{|\mathbf{k'}|-|\mathbf{k}|}.
\end{eqnarray}

After straightforward computation, we finally have
\begin{equation}
\mathrm{Im}\Sigma_{\mathrm{T}}(T) \approx
-\frac{\sqrt[3]{2}e^{\frac{4}{3}}T}{12\sqrt{3}\pi^{\frac{2}{3}}N^{\frac{1}{3}}
\mu^{\frac{2}{3}}} \int_{0}^{\frac{T}{\mu}}d\delta'
\frac{1}{\delta'^{\frac{4}{3}}},
\end{equation}
which is divergent. It is interesting to note that this divergence
is very similar to that appearing in the non-relativistic
fermion-gauge problem (see paper of Lee and Nagaosa in Ref.
\cite{Lee89}). The longitudinal contribution to fermion damping rate
at finite temperature is found to behave as
\begin{equation}
\mathrm{Im}\Sigma_{\mathrm{L}}(T) \propto
\frac{T^2}{\mu}\ln\left(\frac{T}{\mu}\right),
\end{equation}
which is the typical behavior of normal Fermi liquid in two spatial
dimensions. Apparently, the total fermion damping rate is divergent.

\section{Summary and discussion}

In summary, we studied the effect of finite chemical potential $\mu$
on the damping rate of massless Dirac fermions in QED$_3$. At zero
temperature, the total damping rate behaves as
$\mathrm{Im}\Sigma(\varepsilon,\mu)\propto \mu^{-1/3}
\varepsilon^{2/3}$, which vanishes slower than $\varepsilon$ does
near the Fermi surface. This non-Fermi liquid behavior is primarily
generated by the long-range transverse gauge interaction, while the
longitudinal gauge interaction becomes short-ranged and thus only
leads to normal Fermi liquid behavior. It is important to note that
the expression of $\mathrm{Im}\Sigma(\varepsilon)$ at $\mu = 0$ can
not be obtained by simply taking the $\mu \rightarrow 0$ limit from
$\mathrm{Im}\Sigma(\varepsilon,\mu)$ at finite $\mu$. This indicates
that the fermion damping rate displays different
$\varepsilon$-dependence at zero and finite chemical potential,
although Fermi liquid theory breaks down in both cases.

At high fermion density, the Fermi surface becomes very large. Now
the massless Dirac fermion with linear energy spectrum is no longer
a good description for the low-energy excitations. The system is
then described by the non-relativistic fermion-gauge theory
\cite{Lee89, Blok93, GanWong, Polchinski}. Therefore, the results
obtained in this paper are valid only when $\mu$ is not too large.

We have to admit that it is unclear how to get a physically
meaningful fermion damping rate at finite temperature and finite
chemical potential. When $T \ll \mu$, although the longitudinal
component of damping rate has a normal Fermi liquid result, the
transverse component $\mathrm{Im}\Sigma_{\mathrm{T}}(T)$ is
divergent. At present, there seems to be no efficient way to cure
such divergence \cite{Lee89, BellacBook}. In principle, it is
possible to get a divergence-free damping rate by studying the
self-consistent, Eliashberg type, equations of fermion self-energy
function and polarization functions at finite temperature, as we
have done previously \cite{WangLiu}. However, unlike in the case of
zero chemical potential \cite{WangLiu}, we found it difficult to
obtain satisfactory results from the corresponding Eliashberg
equations at finite chemical potential. These problems surely
deserve more thorough investigations in the future.

Finaly, we also calculated the fermion damping rate when the QED$_3$
action has no Maxwell term for the gauge field. A divergence appears
once the Maxwell term is dropped. This divergence has different
origin with that appearing in the damping rate at finite temperature
and finite chemical potential, and arises due to the absence of
Maxwell term. Its appearance may not be surprising since we have
already met it when studying the fermion damping rate at zero
chemical potential \cite{WangLiu}. Unfortunately, we are not aware
of any available method to eliminate the divergence brought by the
absence of Maxwell term at both zero and finite chemical potential.

\section{Acknowledgments}

We thank Dr. W. Li and Dr. F. Xu for discussions. This work is
partly supported by the National Science Foundation of China under
Grant No. 10674122. G.Z.L. also acknowledges the financial support
from the Project Sponsored by the Overseas Academic Training Funds
(OATF), University of Science and Technology of China (USTC).

\appendix

\section{Polarization function at $T = 0$}

When calculating the fermion damping rate at zero temperature $T =
0$ in Sec. IV, it is necessary to first know the temporal and
transverse component of vacuum polarization functions. At $T = 0$,
the integration over parameter $x$ can be carried out analytically,
with the results being presented below. Here the chemical potential
can be taken to be positive: $\mu > 0$.

\subsection{The expression for $\mathrm{Im}\Pi_{00}(\omega,|\mathbf{q}|)$}

We first present the expressions for the region
$|\omega|>|\mathbf{q}|$. For
$0<\mu<\frac{\left|\omega\right|-|\mathbf{q}|}{2}$,
\begin{equation}
\mathrm{Im}\Pi_{00}\left(\omega,|\mathbf{q}|\right)
=\mathrm{sgn}\left(\omega\right) \frac{Ne^2}{16
}\frac{|\mathbf{q}|^2} {\sqrt{\omega^2-|\mathbf{q}|^2} }.
\end{equation}
For $\frac{|\omega|-|\mathbf{q}|}{2} < \mu <
\frac{|\omega|+|\mathbf{q}|}{2}$,
\begin{eqnarray}
\mathrm{Im}\Pi_{00}\left(\omega,|\mathbf{q}|\right) &=&
\mathrm{sgn}(\omega)\frac{Ne^2}{16\pi}
\frac{|\mathbf{q}|^2}{\sqrt{\omega^2-|\mathbf{q}|^2}} \nonumber \\
&&\times\left[\frac{\pi}{2}-A_1\left(\frac{2\mu-|\omega|}
{|\mathbf{q}|}\right)\right],
\end{eqnarray}
where $A_1(y) = y\sqrt{1-y^2}+\arcsin y$. For $\mu
> \frac{|\omega|+|\mathbf{q}|}{2}$,
\begin{equation}
\mathrm{Im}\Pi_{00}(\omega,|\mathbf{q}|)=0.
\end{equation}
We then present the expressions for the region
$|\omega|<|\mathbf{q}|$. For
$0<\mu<\frac{|\mathbf{q}|-|\omega|}{2}$,
\begin{equation}
\mathrm{Im}\Pi_{00}(\omega,|\mathbf{q}|)=0.
\end{equation}
For $\frac{|\mathbf{q}|-|\omega|}{2}
<\mu<\frac{|\mathbf{q}|+|\omega|}{2}$,
\begin{eqnarray}
\mathrm{Im}\Pi_{00}\left(\omega,|\mathbf{q}|\right) &=&
\mathrm{sgn}(\omega)\frac{Ne^2}{16\pi}\frac{|\mathbf{q}|^2}
{\sqrt{|\mathbf{q}|^2-\omega^2}} \nonumber \\
&&\times B_1\left(\frac{2\mu+|\omega|}{|\mathbf{q}|}\right),
\end{eqnarray}
where $B_1\left(y\right)=y\sqrt{y^2-1}-\ln\left|y+\sqrt{y^2-1}\right|$.
For $\mu>\frac{|\mathbf{q}|+|\omega|}{2}$,
\begin{eqnarray}
\mathrm{Im}\Pi_{00}\left(\omega,|\mathbf{q}|\right) &=&
\mathrm{sgn}(\omega)\frac{Ne^2}{16\pi}\frac{|\mathbf{q}|^2}
{\sqrt{|\mathbf{q}|^2-\omega^2}} \nonumber \\
&&\times\left[B_1\left(\frac{2\mu+|\omega|}{|\mathbf{q}|}\right)\right.\nonumber\\
&&\left.-B_1\left(\frac{2\mu-|\omega|}{|\mathbf{q}|}\right)\right].
\end{eqnarray}

\subsection{The expression for
$\mathrm{Re}\Pi_{00}(\omega,|\mathbf{q}|)$}

We first present the expressions for the region
$|\omega|>|\mathbf{q}|$. For
$0<\mu<\frac{|\omega|-|\mathbf{q}|}{2}$,
\begin{eqnarray}
\mathrm{Re}\Pi_{00}\left(\omega,|\mathbf{q}|\right)
&=&\frac{Ne^2\mu}{2\pi} - \frac{Ne^2}{16\pi}
\frac{|\mathbf{q}|^2}{\sqrt{\omega^2-|\mathbf{q}|^2}} \nonumber \\
&&\times\left[B_1\left(\frac{|\omega|+2\mu}{|\mathbf{q}|}\right)\right.\nonumber\\
&&\left.-B_1\left(\frac{|\omega|-2\mu}{|\mathbf{q}|}\right)\right].
\end{eqnarray}
For $\frac{|\omega|-|\mathbf{q}|}{2} < \mu <
\frac{|\omega|+|\mathbf{q}|}{2}$,
\begin{eqnarray}
\mathrm{Re}\Pi_{00}\left(\omega,|\mathbf{q}|\right) &=&
\frac{Ne^2\mu}{2\pi} -
\frac{Ne^2}{16\pi}\frac{|\mathbf{q}|^2}{\sqrt{\omega^2-|\mathbf{q}|^2}}
\nonumber \\
&&\times B_1\left(\frac{|\omega|+2\mu}{|\mathbf{q}|}\right).
\end{eqnarray}
For $\mu>\frac{|\omega|+|\mathbf{q}|}{2}$,
\begin{eqnarray}
\mathrm{Re}\Pi_{00}\left(\omega,|\mathbf{q}|\right) &=&
\frac{Ne^2\mu}{2\pi} -
\frac{Ne^2}{16\pi}\frac{|\mathbf{q}|^2}{\sqrt{\omega^2-|\mathbf{q}|^2}}
\nonumber \\
&& \times \left[B_1\left(\frac{2\mu+|\omega|}{|\mathbf{q}|}\right)\right.\nonumber\\
&&\left.-B_1\left(\frac{2\mu-|\omega|}{|\mathbf{q}|}\right)\right].
\end{eqnarray}
We then present the expressions for the region
$|\omega|<|\mathbf{q}|$. For
$0<\mu<\frac{|\mathbf{q}|-|\omega|}{2}$,
\begin{eqnarray}
\mathrm{Re}\Pi_{00}\left(\omega,|\mathbf{q}|\right) &=&
\frac{Ne^2\mu}{2\pi}+
\frac{Ne^2}{16\pi}\frac{|\mathbf{q}|^2}{\sqrt{|\mathbf{q}|^2-\omega^2}}\nonumber
\\
&& \times
\left[\pi-A_1\left(\frac{2\mu+|\omega|}{|\mathbf{q}|}\right)\right.
\nonumber\\
&&\left.-A_1\left(\frac{2\mu-|\omega|}{|\mathbf{q}|}\right)\right].
\end{eqnarray}
For $\frac{|\mathbf{q}|-|\omega|}{2}<\mu<
\frac{|\mathbf{q}|+|\omega|}{2}$,
\begin{eqnarray}
\mathrm{Re}\Pi_{00}\left(\omega,|\mathbf{q}|\right)
&=&\frac{Ne^2\mu}{2\pi}+
\frac{Ne^2}{16\pi}\frac{|\mathbf{q}|^2}{\sqrt{|\mathbf{q}|^2-\omega^2}}
\nonumber \\
&& \times \left[\frac{\pi}{2} -
A_1\left(\frac{2\mu-|\omega|}{|\mathbf{q}|}\right)\right].
\end{eqnarray}
For $\mu>\frac{|\mathbf{q}|+|\omega|}{2}$,
\begin{equation}
\mathrm{Re}\Pi_{00}\left(\omega,|\mathbf{q}|\right) =
\frac{Ne^2\mu}{2\pi}.
\end{equation}

\subsection{The expression for
$\mathrm{Im}\Pi_{\bot}\left(\omega,|\mathbf{q}|\right)$}

We first present the expressions for the region
$|\omega|>|\mathbf{q}|$. For
$0<\mu<\frac{|\omega|-|\mathbf{q}|}{2}$,
\begin{equation}
\mathrm{Im}\Pi_{\bot}\left(\omega,|\mathbf{q}|\right)
=-\mathrm{sgn}\left(\omega\right)
\frac{Ne^2}{16}\sqrt{\omega^2-|\mathbf{q}|^2}.
\end{equation}
For $\frac{|\omega|-|\mathbf{q}|}{2}
<\mu<\frac{|\omega|+|\mathbf{q}|}{2}$,
\begin{eqnarray}
\mathrm{Im}\Pi_{\bot}\left(\omega,|\mathbf{q}|\right)
&=&-\mathrm{sgn}\left(\omega\right)
\frac{Ne^2}{16\pi}\sqrt{\omega^2-|\mathbf{q}|^2}\nonumber
\\
&&\times\left[\frac{\pi}{2}-A_2\left(\frac{2\mu-|\omega|}{|\mathbf{q}|}\right)
\right],
\end{eqnarray}
where $A_2\left(y\right)=-y\sqrt{1-y^2}+\arcsin y$.
For $\mu>\frac{|\omega|+|\mathbf{q}|}{2}$,
\begin{equation}
\mathrm{Im}\Pi_{\bot}\left(\omega,|\mathbf{q}|\right)=0.
\end{equation}
We then present the expressions for the region
$|\omega|<|\mathbf{q}|$. For
$0<\mu<\frac{|\mathbf{q}|-|\omega|}{2}$,
\begin{equation}
\mathrm{Im}\Pi_{\bot}\left(\omega,|\mathbf{q}|\right)=0.
\end{equation}
For $\frac{|\mathbf{q}|-|\omega|}{2}
<\mu<\frac{|\mathbf{q}|+|\omega|}{2}$,
\begin{eqnarray}
\mathrm{Im}\Pi_{\bot}\left(\omega,|\mathbf{q}|\right)
&=&-\mathrm{sgn}\left(\omega\right)
\frac{Ne^2}{16\pi}\sqrt{|\mathbf{q}|^2-\omega^2}\nonumber
\\
&&\times B_2\left(\frac{2\mu+|\omega|}{|\mathbf{q}|}\right),
\end{eqnarray}
where $B_2\left(y\right)=y\sqrt{y^2-1}+\ln\left|y+\sqrt{y^2-1}\right|$.
For $\mu>\frac{|\mathbf{q}|+|\omega|}{2}$,
\begin{eqnarray}
\mathrm{Im}\Pi_{\bot}\left(\omega,|\mathbf{q}|\right)
&=&-\mathrm{sgn}\left(\omega\right)
\frac{Ne^2}{16\pi}\sqrt{|\mathbf{q}|^2-\omega^2}\nonumber
\\
&&\times\left[B_2\left(\frac{2\mu+|\omega|}{|\mathbf{q}|}\right)\right.\nonumber\\
&&\left.-B_2\left(\frac{2\mu-|\omega|}{|\mathbf{q}|}\right)\right].
\end{eqnarray}

\subsection{The expression for
$\mathrm{Re}\Pi_{\bot}\left(\omega,|\mathbf{q}|\right)$}

We first present the expressions for the region
$|\omega|>|\mathbf{q}|$. For
$0<\mu<\frac{|\omega|-|\mathbf{q}|}{2}$,
\begin{eqnarray}
\mathrm{Re}\Pi_{\bot}\left(\omega,|\mathbf{q}|\right)
&=&\frac{Ne^2\mu}{2\pi}\frac{\omega^2}{|\mathbf{q}|^2}
-\frac{Ne^2}{16\pi}\sqrt{\omega^2-|\mathbf{q}|^2}\nonumber
\\
&&\times\left[B_2\left(\frac{|\omega|+2\mu}{|\mathbf{q}|}\right)\right.\nonumber\\
&&\left.-B_2\left(\frac{|\omega|-2\mu}{|\mathbf{q}|}\right)\right].
\end{eqnarray}
For $\frac{|\omega|-|\mathbf{q}|}{2}
<\mu<\frac{|\omega|+|\mathbf{q}|}{2}$,
\begin{eqnarray}
\mathrm{Re}\Pi_{\bot}\left(\omega,|\mathbf{q}|\right)
&=&\frac{Ne^2\mu}{2\pi}\frac{\omega^2}{|\mathbf{q}|^2}
-\frac{Ne^2}{16\pi}\sqrt{\omega^2-|\mathbf{q}|^2}\nonumber
\\
&&\times\left[B_2\left(\frac{|\omega|+2\mu}{|\mathbf{q}|}\right)\right].
\end{eqnarray}
For $\mu>\frac{|\omega|+|\mathbf{q}|}{2}$,
\begin{eqnarray}
\mathrm{Re}\Pi_{\bot}\left(\omega,|\mathbf{q}|\right)
&=&\frac{Ne^2\mu}{2\pi}\frac{\omega^2}{|\mathbf{q}|^2}
-\frac{Ne^2}{16\pi}\sqrt{\omega^2-|\mathbf{q}|^2}\nonumber
\\
&&\times\left[B_2\left(\frac{2\mu+|\omega|}{|\mathbf{q}|}\right)\right.\nonumber\\
&&\left.-B_2\left(\frac{2\mu-|\omega|}{|\mathbf{q}|}\right)\right].
\end{eqnarray}
We then present the expressions for the region
$|\omega|<|\mathbf{q}|$. For
$0<\mu<\frac{|\mathbf{q}|-|\omega|}{2}$,
\begin{eqnarray}
\mathrm{Re}\Pi_{\bot}\left(\omega,|\mathbf{q}|\right)
&=&\frac{Ne^2\mu}{2\pi} \frac{\omega^2}{|\mathbf{q}|^2}
+\frac{Ne^2}{16\pi}\sqrt{|\mathbf{q}|^2-\omega^2}\nonumber
\\
&&\times\left[\pi-A_2\left(\frac{2\mu+|\omega|}{|\mathbf{q}|}\right)\right.\nonumber\\
&&\left.-A_2\left(\frac{2\mu-|\omega|}{|\mathbf{q}|}\right)\right].
\end{eqnarray}
For $\frac{|\mathbf{q}|-|\omega|}{2}
<\mu<\frac{|\mathbf{q}|+|\omega|}{2}$,
\begin{eqnarray}
\mathrm{Re}\Pi_{\bot}\left(\omega,|\mathbf{q}|\right)
&=&\frac{Ne^2\mu}{2\pi} \frac{\omega^2}{|\mathbf{q}|^2}
+\frac{Ne^2}{16\pi}\sqrt{|\mathbf{q}|^2-\omega^2}\nonumber
\\
&&\times\left[\frac{\pi}{2}-A_2\left(\frac{2\mu-|\omega|}{|\mathbf{q}|}
\right)\right].
\end{eqnarray}
For $\mu>\frac{|\mathbf{q}|+|\omega|}{2}$,
\begin{equation}
\mathrm{Re}\Pi_{\bot}\left(\omega,|\mathbf{q}|\right)
=\frac{Ne^2\mu}{2\pi} \frac{\omega^2}{|\mathbf{q}|^2}.
\end{equation}


\begin{thebibliography}{99}


\bibitem{Holstein}
T. Holstein, R. E. Norton, and P. Pincus, Phys. Rev. B {\bf 8}, 2649
(1973).

\bibitem{Varma2002}
C. M. Varma, Z. Nussinov, and W. van Saarloos, Phys. Rep. {\bf 361},
267 (2002) .

\bibitem{Reizer89}
M. Y. Reizer, Phys. Rev. B {\bf 39}, 1602 (1989).

\bibitem{Lee89}
P. A. Lee, Phys. Rev. Lett. {\bf 63}, 680 (1989); P. A. Lee and N.
Nagaosa, Phys. Rev. B {\bf 46}, 5621 (1992).

\bibitem{Blok93}
B. Blok and H. Monien, Phys. Rev. B {\bf 47}, 3454 (1993); D. V.
Khveshchenko, R. Hlubina, and T. M. Rice, Phys. Rev. B {\bf 48},
10766 (1993); B. L. Altshuler, L. B. Ioffe, and A. J. Millis, Phys.
Rev. B {\bf 50}, 14048 (1994).

\bibitem{GanWong}
J. Gan and E. Wong, Phys. Rev. Lett. {\bf 71}, 4226 (1993).

\bibitem{Polchinski}
J. Polchinski, Nucl. Phys. B {\bf 422}, 617 (1994).

\bibitem{Lee05}
P. A. Lee, N. Nagaosa, and X.-G. Wen, Rev. Mod. Phys. {\bf 78}, 17
(2006).

\bibitem{Pisarski}
R. D. Pisarski, Phys. Rev. Lett. {\bf 63}, 1129 (1989); Phys. Rev. D
{\bf 47}, 5589 (1993).

\bibitem{Blaizot}
J.-P. Blaizot and E. Iancu, Phys. Rev. Lett. {\bf 76}, 3080 (1996);
Phys. Rev. D {\bf 55}, 973 (1997).

\bibitem{Lebellac}
M. Le Bellac and C. Manuel, Phys. Rev. D {\bf 55}, 3215 (1997).

\bibitem{Castro}
A. H. Castro Neto, F. Guinea, N. M. R. Peres, K. S. Novoselov, and
A. K. Geim, Rev. Mod. Phys. {\bf 81}, 109 (2009).

\bibitem{Appelquist88}
T. Appelquist, D. Nash, and L. C. R. Wijewardhana, Phys. Rev. Lett.
{\bf 60}, 2575 (1988); D. Nash, Phys. Rev. Lett. {\bf 62}, 3024
(1989); E. Dagotto, J. B. Kogut, and A. Koci\'{c}, Phys. Rev. Lett.
{\bf 62}, 1083 (1989); T. W. Appelquist and L. C. R. Wijewardhana,
arXiv:hep-ph/0403250v4, (2004).

\bibitem{Maris95}
P. Maris, Phys. Rev. D {\bf 52}, 6087 (1995).

\bibitem{Affleck}
I. Affleck and J. B. Marston, Phys. Rev. B {\bf 37}, 3774 (1988); L.
B. Ioffe and A. I. Larkin, Phys. Rev. B {\bf 39}, 8988 (1989).

\bibitem{Kim97}
D. H. Kim, P. A. Lee, and X.-G. Wen, Phys. Rev. Lett. {\bf 79}, 2109
(1997); W. Rantner and X.-G. Wen, Phys. Rev. Lett. {\bf 86}, 3871
(2001).

\bibitem{Franz}
M. Franz and Z. Te\v{s}anovi\'{c}, Phys. Rev. Lett. {\bf 87}, 257003
(2001); I. F. Herbut, Phys. Rev. B {\bf 66}, 094504 (2002).

\bibitem{Kim99}
J. B. Marston, Phys. Rev. Lett. {\bf 64}, 1166 (1990); D. H. Kim and
P. A. Lee, Ann. Phys. (N.Y.) {\bf 272}, 130 (1999); I. F. Herbut,
Phys. Rev. Lett. {\bf 88}, 047006 (2002); Z. Te\v{s}anovi\'{c}, O.
Vafek, and M. Franz, Phys. Rev. B {\bf 65}, 180511 (2002); G.-Z. Liu
and G. Cheng, Phys. Rev. D {\bf 67}, 065010 (2003); G.-Z. Liu, Phys.
Rev. B {\bf 71}, 172501 (2005).

\bibitem{Ran07}
Y. Ran, M. Hermele, P. A. Lee, and X.-G. Wen, Phys. Rev. Lett. {\bf
98}, 117205 (2007).

\bibitem{WangLiu}
J.-R. Wang and G.-Z. Liu, Nucl. Phys. B {\bf 832}, 441 (2010).

\bibitem{Sachdev}
S. Sachdev, arXiv:0910.1139v1.

\bibitem{Giuliani}
G. F. Giuliani and G. Vignale, \emph{Quantum} \emph{Theory}
\emph{of} \emph{the} \emph{Elentron} \emph{Liquid} (Cambridge
University Press, Cambridge, 2005).

\bibitem{Franz02}
M. Franz, Z. Tesanovic, O. Vafek, Phys. Rev. B {\bf 66}, 054535
(2002).

\bibitem{Rech}
A. V. Chubukov, C. Pepin, and J. Rech, Phys. Rev. Lett. {\bf 92},
147003 (2004); J. Rech, C. Pepin, and A. V. Chubukov, Phys. Rev. B
{\bf 74}, 195126 (2006).

\bibitem{Sunkai}
K. Sun, B. M. Fregoso, M. J. Lawler, and E. Fradkin, Phys. Rev. B
{\bf 78}, 085124 (2008).

\bibitem{Metlitski}
M. A. Metlitski and S. Sachdev, arXiv:1001.1153v3.

\bibitem{Metzner}
W. Metzner, D. Rohe, and S. Andergassen, Phys. Rev. Lett. {\bf 91},
066402 (2003); L. Dell¡¯Anna and W. Metzner, Phys. Rev. B {\bf 73},
045127 (2006).

\bibitem{BellacBook}
M. Le Bellac, \emph{Thermal Field Theory} (Cambridge University
Press, Cambridge, 1996)


\end{thebibliography}
\end{document}